\documentclass{llncs}

\usepackage{amssymb}
\usepackage{amsmath}
\usepackage{graphicx}
\usepackage{braket}

\begin{document}
\bibliographystyle{splncs}

\title{Limit theorem for a time-dependent \\coined quantum walk on the line}
\author{Takuya Machida\inst{1} \and Norio Konno\inst{2}}
\institute{Department of Applied Mathematics, Faculty of Engineering,\\Yokohama National University,
Hodogaya, Yokohama, 240-8501, Japan,\\
\email{bunchin@meiji.ac.jp}
\and
\email{konno@ynu.ac.jp}}

\maketitle

\abstract{We study time-dependent discrete-time quantum walks on the one-dimensional lattice.
We compute the limit distribution of a two-period quantum walk defined by two orthogonal matrices.
For the symmetric case, the distribution is determined by one of two matrices.
Moreover, limit theorems for two special cases are presented.}

\section{Introduction}

The discrete-time quantum walk (QW) was first intensively studied by Ambainis {\it et al.} \cite{ambainis}.
The QW is considered as a quantum generalization of the classical random walk.
The random walker in position $x\in\mathbb{Z}=\left\{\,0,\pm 1,\pm 2,\ldots\,\right\}$ at time
$t\left(\in \left\{\,0,1,2,\ldots\,\right\}\right)$ moves to $x-1$ at time $t+1$ with
probability $p$, or $x+1$ with probability $q\,(\,=1-p)$.
In contrast, the evolution of the quantum walker is defined by replacing $p$ and $q$ with $2\times 2$ matrices $P$ and $Q$, respectively.
Note that $U=P+Q$ is a unitary matrix.
A main difference between the classical walk and the QW is seen on the particle spreading.
Let $\sigma(t)$ be the standard deviation of the walk at time $t$.
That is, $\sigma(t)=\sqrt{\mathbb{E}(X_t^2)-\mathbb{E}(X_t)^2}$, where $X_t$ is the position of the quantum walker at time $t$ and $\mathbb{E}(Y)$ denotes the expected value of $Y$.
Then the classical case is a diffusive behavior, $\sigma(t)\sim \sqrt{t}$, while the quantum case is ballistic, $\sigma(t)\sim t$ (see \cite{ambainis}, for example).

In the context of quantum computation, the QW is applied to several quantum algorithms.
By using the quantum algorithm, we solve a problem quadratically faster than the corresponding classical algorithm.
As a well-known quantum search algorithm, Grover's algorithm was presented.
The algorithm solves the following problem: in a search space of $N$ vertices, one can find a marked vertex.
The corresponding classical search requires $O(N)$ queries.
However, the search needs only $O(\sqrt{N})$ queries.
As well as the Grover algorithm, the QW can also search a marked vertex with a quadratic speed up, see Shenvi {\it et al.} \cite{shenvi}.
It has been reported that quantum walks on regular graphs (e.g., lattice, hypercube, complete graph) give faster searching than classical walks.
The Grover search algorithm can also be interpreted as a QW on complete graph.
Decoherence is an important concept in quantum information processing.
In fact, decoherence on QWs has been extensively investigated, see Kendon \cite{kendon}, for example.
However, we should note that our results are not related to the decoherence in QWs.
Physically, Oka {\it et al.} \cite{oka} pointed out that the Landau-Zener transition dynamics can be mapped to a QW and showed the localization of the wave functions.

In the present paper, we consider the QW whose dynamics is determined by a sequence of time-dependent matrices, $\left\{U_t:t=0,1,\ldots\right\}$.
Ribeiro {\it et al.} \cite{ribeiro} numerically showed that periodic sequence is ballistic, random sequence is diffusive, and Fibonacci sequence is sub-ballistic.
Mackay {\it et al.} \cite{mackay} and Ribeiro {\it et al.} \cite{ribeiro} investigated some random sequences and reported that the probability distribution of the QW converges to a binomial distribution by averaging over many trials by numerical simulations.
Konno \cite{konno_2005_2} proved their results by using a path counting method.
By comparing with a position-dependent QW introduced by W\'{o}jcik {\it et al.} \cite{wojcik}, Ba\~{n}uls {\it et al.} \cite{banuls} discussed a dynamical localization of the corresponding time-dependent QW.

In this paper, we present the
weak limit theorem for the two-period time-dependent QW whose unitary matrix $U_t$ is
an orthogonal matrix.
Our approach is based on the Fourier transform method introduced by Grimmett {\it et al.} \cite{grimmett}.
We think that it would be difficult to calculate the limit distribution for the general $n$-period ($n=3,4,\ldots$) walk.
However, we find out a class of time-dependent QWs whose limit probability distributions result in that of the usual (i.e., one-period) QW.
As for the position-dependent QW, a similar result can be found in Konno \cite{konno_2009}.

The present paper is organized as follows.
In Sect. 2, we define the time-dependent QW.
Section 3 treats the two-period time-dependent QW.
By using the Fourier transform, we obtain the limit distribution.
Finally, in Sect. 4, we consider two special cases of time-dependent QWs.
We show that the limit distribution of the walk is the same as that of the usual one.

\section{Time-dependent QW}

In this section we define the time-dependent QWs.
Let $\ket{x}$ ($x\in\mathbb{Z}$) be infinite components vector which denotes the position of the walker.
Here,  $x$-th component of $\ket{x}$ is 1 and the other is 0.
Let $\ket{\psi_{t}(x)} \in \mathbb{C}^2$ be the amplitude of the walker in position $x$ at time $t$, where $\mathbb{C}$ is the set of complex numbers.
The time-dependent QW at time $t$ is expressed by
\begin{equation}
 \ket{\Psi_t}=\sum_{x\in\mathbb{Z}}\ket{x}\otimes\ket{\psi_{t}(x)}.
\end{equation}
To define the time evolution of the walker, we introduce a unitary matrix
\begin{equation}
 U_t=\left[\begin{array}{cc}
      a_t& b_t\\ c_t&d_t
	   \end{array}\right],
\end{equation}
where $a_t, b_t, c_t, d_t\,\in\mathbb{C}$ and $a_tb_tc_td_t\neq 0\,\,(t=0,1,\ldots)$.
Then $U_t$ is divided into $P_t$ and $Q_t$ as follows:
\begin{equation}
 P_t=\left[\begin{array}{cc}
      a_t& b_t\\ 0&0
	   \end{array}\right],\,
 Q_t=\left[\begin{array}{cc}
      0&0 \\ c_t&d_t
	   \end{array}\right].
\end{equation}

The evolution is determined by
\begin{equation}
 \ket{\Psi_{t+1}}
 =\sum_{x\in\mathbb{Z}}\ket{x}\otimes\left(P_t\ket{\psi_t(x+1)}+Q_t\ket{\psi_t(x-1)}\right).
\end{equation}
Let $||\ket{y}||^2=\braket{y|y}$.
The probability that the quantum walker $X_t$ is in position $x$ at time $t$, $P(X_t=x)$, is defined by
\begin{equation}
 P(X_t=x)=||\ket{\psi_t(x)}||^2.
\end{equation}
Moreover, the Fourier transform $\ket{\hat{\Psi}_{t}(k)}\,(k\in\left[0,2\pi\right))$ is given by

\begin{equation}
 \ket{\hat{\Psi}_{t}(k)}=\sum_{x\in\mathbb{Z}} e^{-ikx}\ket{\psi_t(x)},
\end{equation}
with $i=\sqrt{-1}$.
By the inverse Fourier transform, we have
\begin{equation}
 \ket{\psi_t(x)}=\int_{0}^{2\pi}\frac{dk}{2\pi}e^{ikx}\ket{\hat\Psi_{t}(k)}.
\end{equation}
The time evolution of $\ket{\hat{\Psi}_{t}(k)}$ is
\begin{equation}
 \ket{\hat{\Psi}_{t+1}(k)}=\hat U_t(k)\ket{\hat{\Psi}_{t}(k)},\label{eq:timeevo}
\end{equation}
where $\hat U_{t}(k)=R(k)U_t$ and
$R(k)=\left[\begin{array}{cc}
       e^{ik}&0 \\0&e^{-ik}
	    \end{array}\right]$.
We should remark that $R(k)$ satisfies $R(k_1)R(k_2)=R(k_1+k_2)$ and $R(k)^{\ast}=R(-k)$, where $\ast$ denotes the conjugate transposed operator.
From (\ref{eq:timeevo}), we see that
\begin{equation}
 \ket{\hat{\Psi}_{t}(k)}=\hat U_{t-1}(k)\hat U_{t-2}(k)\cdots\hat U_0(k)\ket{\hat\Psi_{0}(k)}.
\end{equation}
Note that, when $U_t=U$ for any $t$, the walk
becomes a usual one-period walk, and $\ket{\hat{\Psi}_{t}(k)}=\hat
U(k)^t\ket{\hat\Psi_{0}(k)}$.
Then the probability distribution of the usual walk is
\begin{equation}
 P(X_t=x)=\left|\left|\int_{0}^{2\pi}\frac{dk}{2\pi}e^{ikx}\hat U(k)^t\ket{\hat\Psi_{0}(k)}\right|\right|^2.
\end{equation}
In Sect. 4, we will use this relation.
In the present paper, we take the initial state as
\begin{equation}
 \ket{\psi_0(x)}=\left\{\begin{array}{ll}
		 \!{}^T[\,\alpha, \,\beta\,]& (x=0)\\[2mm]
			\!{}^T[\,0,\,0\,]& (x\neq 0)
		       \end{array}\right.,
\end{equation}
where $|\alpha|^2+|\beta|^2=1$ and $T$ is the transposed operator.
We should note that $\ket{\hat\Psi_{0}(k)}=\ket{\psi_0(0)}$.
\clearpage
\section{Two-period QW}

In this section we consider the two-period QW and calculate the limit distribution.
We assume that $\left\{U_t:t=0,1,\ldots\right\}$ is a sequence of orthogonal matrices with $U_{2s}=H_0$ and $U_{2s+1}=H_1$ $(s=0,1,\ldots)$, where
\begin{equation}
 H_0=\left[\begin{array}{cc}
      a_0&b_0 \\c_0& d_0
	   \end{array}\right],\,
	   H_1=\left[\begin{array}{cc}
		a_1&b_1 \\c_1& d_1
		     \end{array}\right].
\end{equation}
Let
\begin{equation}
 f_{K}(x;a)=\frac{\sqrt{1-|a|^2}}{\pi(1-x^2)\sqrt{|a|^2-x^2}} \,I_{(-|a|,|a|)}(x),
\end{equation}
where $I_{A}(x)=1$ if $x\in A$, $I_{A}(x)=0$ if $x\notin A$.
Then we obtain the following main result of this paper:

\vspace{5mm}

\begin{theorem}
\begin{equation}
 \frac{X_{t}}{t}\,\Rightarrow\,Z,
\end{equation}
where $\Rightarrow$ means the weak convergence (i.e., the convergence of the distribution) and $Z$ has the density function $f(x)$ as follows:

\begin{enumerate}
\renewcommand{\labelenumi}{(\roman{enumi})}
 \item If $\det (H_1H_0) > 0$, then\\

       \begin{equation}
       f(x)=f_{K}(x;a_{\xi})
       \left[1-\left\{\left|\alpha\right|^2-\left|\beta\right|^2
       +\frac{\left(\alpha\overline{\beta}+\overline{\alpha}\beta\right)
       b_0}{a_0}\right\}x\right],
       \end{equation}
       where $|a_{\xi}|=\min\left\{|a_0|,\,|a_1|\right\}$.\vspace{5mm}

 \item If $\det(H_1H_0)<0$, then\\

       \begin{equation}
       f(x)=f_{K}(x;a_0a_1)
       \left[1-\left\{\left|\alpha\right|^2-\left|\beta\right|^2
       +\frac{\left(\alpha\overline{\beta}+\overline{\alpha}\beta\right)
       b_0}{a_0}\right\}x\right].
       \end{equation}

\end{enumerate}
\end{theorem}
\vspace{5mm}

\begin{figure}[h]
 \begin{center}
 \begin{minipage}{40mm}
  \begin{center}
   \includegraphics[scale=0.4]{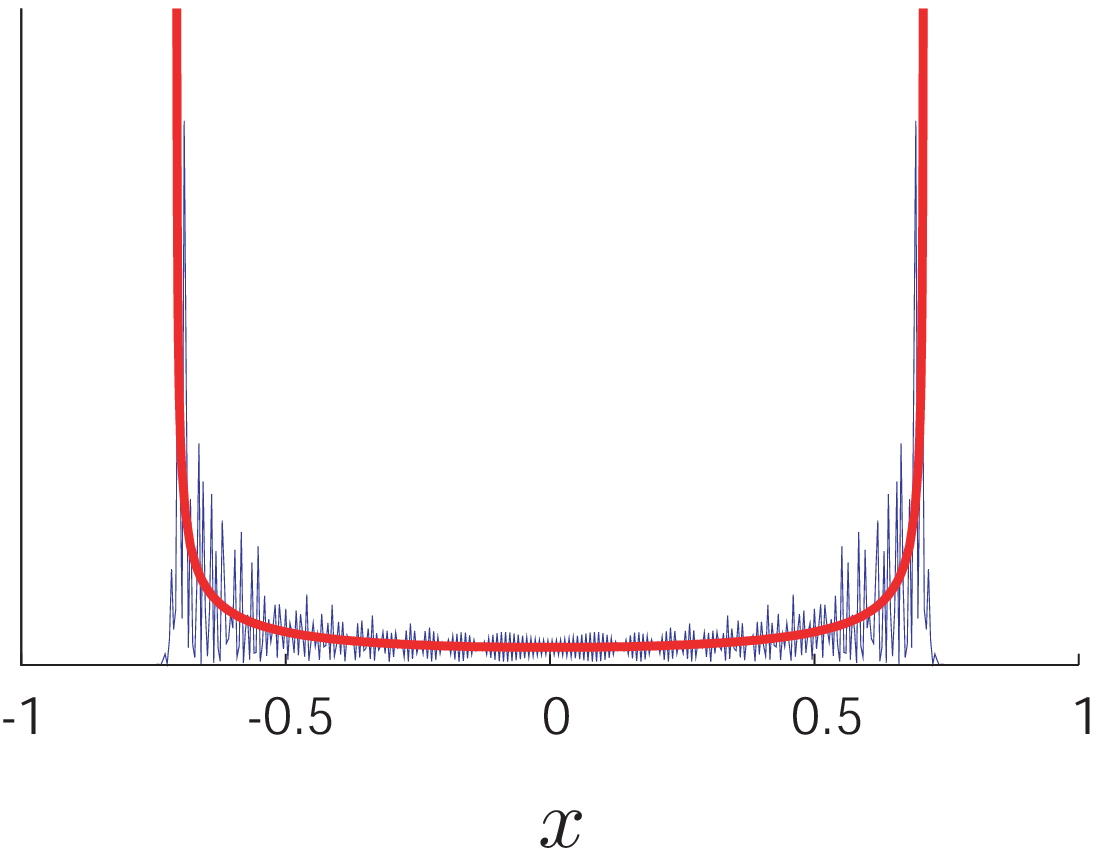}\\
   {(a) $a_0=\cos(\pi/4)$,\\\quad\,\,\,$a_1=\cos(\pi/6)$}
  \end{center}
 \end{minipage}\hspace{1cm}
 \begin{minipage}{40mm}
  \begin{center}
   \includegraphics[scale=0.4]{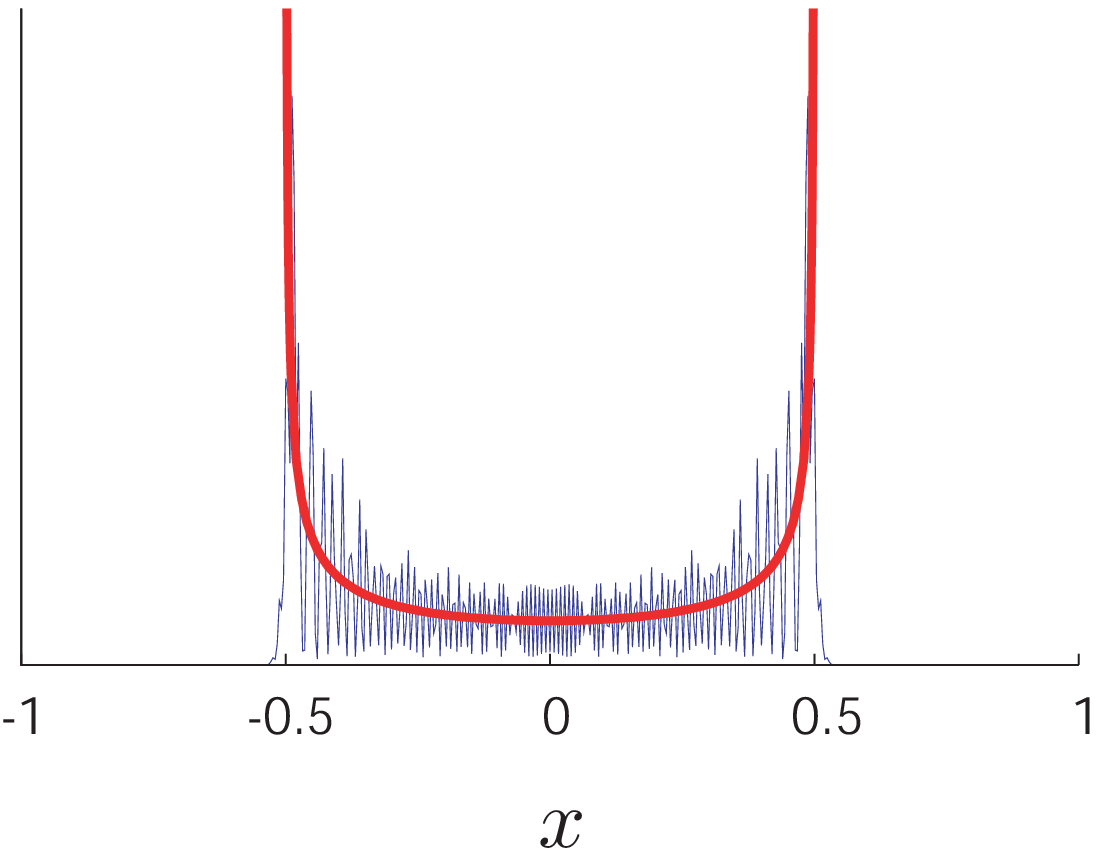}\\
   {(b) $a_0=\cos(\pi/4)$,\\\quad\,\,\,$a_1=\cos(\pi/3)$}
  \end{center}
 \end{minipage}
 \caption{The limit density function $f(x)$ (thick line) and the probability distribution at time $t=500$ (thin line).}
  \label{fig:density}
 \end{center}
\end{figure}

If the two-period walk with $\det(H_1H_0)>0$ has a symmetric distribution, then the density of $Z$ becomes $f_{K}(x;a_{\xi})$.
That is, $Z$ is determined by either $H_0$ or $H_1$.
Figure \ref{fig:density} (a) shows that the limit density of the two-period QW for $a_0=\cos(\pi/4)$ and $a_1=\cos(\pi/6)$ is the same as that for the usual (one-period) QW for $a_0$, since $\left|a_0\right|<\left|a_1\right|$. Similarly, Fig. \ref{fig:density} (b) shows that the limit density of the two-period QW for $a_0=\cos(\pi/4)$ and $a_1=\cos(\pi/3)$ is equivalent to that for the usual (one-period) QW for $a_1$, since $\left|a_0\right|>\left|a_1\right|$.
\clearpage

\begin{proof}

Our approach is due to Grimmett {\it et al.} \cite{grimmett}.
The Fourier transform becomes
\begin{equation}
 \ket{\hat{\Psi}_{2t}(k)}=\left(\hat{H}_1(k)\hat{H}_0(k)\right)^t\ket{\hat\Psi_{0}(k)},
\end{equation}
where $\hat{H}_{\gamma}(k)=R(k)H_{\gamma}\,(\gamma=0,1)$.
We assume
\begin{equation}
 H_{\gamma}=\left[\begin{array}{cc}
      \cos\theta_{\gamma}&\sin\theta_{\gamma} \\ \sin\theta_{\gamma}&-\cos\theta_{\gamma}
	   \end{array}\right],
\end{equation}
with $\theta_{\gamma} \neq \frac{\pi n}{2}\,(n\in\mathbb{Z})$ and $\theta_0\neq \theta_1$.
For the other case, the argument is nearly identical to this case, so we will omit it.
The two eigenvalues $\lambda_j(k)\,(j=0,1)$ of
$\hat{H}_1(k)\hat{H}_0(k)$ are given by
\begin{equation}
 \lambda_j(k)=c_1c_2\cos 2k+s_1s_2+(-1)^j \,i\sqrt{1-(c_1c_2\cos 2k+s_1s_2)^2},
\end{equation}
where $c_{\gamma}=\cos\theta_{\gamma},\,s_{\gamma}=\sin\theta_{\gamma}$.
The eigenvector $\ket{v_j(k)}$ corresponding to $\lambda_j(k)$ is
\begin{equation}
 \ket{v_j(k)}=\left[\begin{array}{c}
	       s_1c_2e^{2ik}-c_1s_2\\
		     \left\{-c_1c_2\sin 2k+(-1)^j \sqrt{1-(c_1c_2\cos 2k+s_1s_2)^2}\right\}i
		    \end{array}\right].
\end{equation}
The Fourier transform $\ket{\hat\Psi_{0}(k)}$ is expressed by normalized
eigenvectors $\ket{v_j(k)}$ as follows:
\begin{equation}
 \ket{\hat\Psi_{0}(k)}=\sum_{j=0}^{1}\braket{v_j(k)|\hat\Psi_{0}(k)}\ket{v_j(k)}.
\end{equation}
Therefore we have
\begin{eqnarray}
 \ket{\hat\Psi_{2t}(k)}&=&\left(\hat H_1(k)\hat H_0(k)\right)^t\ket{\hat\Psi_{0}(k)}\nonumber\\
 &=&\sum_{j=0}^{1}\lambda_j(k)^t\braket{v_j(k)|\hat\Psi_{0}(k)}\ket{v_j(k)}.
\end{eqnarray}

The {\it r}-th moment of $X_{2t}$ is
\begin{eqnarray}
 E((X_{2t})^r)&=&\sum_{x\in \mathbb{Z}}x^r P(X_{2t}=x)\nonumber\\
&=&\int_{0}^{2\pi}\frac{dk}{2\pi}
 \bra{\hat\Psi_{2t}(k)}\left(D^r\ket{\hat\Psi_{2t}(k)}\right)\nonumber\\
 &=&\int_{0}^{2\pi}\sum_{j=0}^{1}(t)_r\lambda_j(k)^{-r}(D\lambda_j(k))^r\left|\braket{v_j(k)|\hat\Psi_{0}(k)}\right|^2\nonumber\\
 &&+O(t^{r-1}),
\end{eqnarray}
where $D=i(d/dk)$ and $(t)_r=t(t-1)\times\cdots\times(t-r+1)$.
Let $h_j(k)=D\lambda_j(k)/2\lambda_j(k)$. Then we obtain
\begin{equation}
 E((X_{2t}/2t)^r) \rightarrow \int_{\Omega_0}\frac{dk}{2\pi}\sum_{j=0}^{1}
 h^r_j(k)|\braket{v_j(k)|\hat\Psi_0(k)}|^2 \quad(\,t\rightarrow\infty).
\end{equation}
Substituting $h_j(k)=x$, we have
\begin{equation}
 \lim_{t\rightarrow\infty}E((X_{2t}/2t)^r)=\int_{-|c_{\xi}|}^{|c_{\xi}|}x^r\,f(x)\,dx ,
\end{equation}
where
\begin{equation}
 f(x)=f_{K}(x;c_\xi)\left[1-\left\{|\alpha|^2-|\beta|^2
 +\frac{(\alpha\bar\beta+\bar\alpha\beta)s_1}{c_1}\right\}x\right],
\end{equation}
and $|c_\xi|=|\cos\theta_\xi|=\min\left\{|\cos\theta_0|,|\cos\theta_1|\right\}$.
Since $f(x)$ is the limit density function, the proof is complete.
\qed
\end{proof}

\section{Special cases in time-dependent QWs}

In the previous section, we have obtained the limit theorem for the two-period QW determined by two orthogonal matrices.
For other two-period case and general $n$-period ($n\geq 3$) case, we think that it would be hard to get the limit theorem in a similar fashion.
Here we consider two special cases in the
time-dependent QWs and give the weak limit theorems.

\subsection{Case 1}

Let us consider the QW whose evolution is determined by the following unitary matrix:
\begin{equation}
 U_t=\left[\begin{array}{cc}
      ae^{iw_t}& b \\ c& de^{-iw_t}
	   \end{array}\right],
\end{equation}
with $a,b,c,d\,\in\,\mathbb{C}$.
Here $w_t\in\mathbb{R}$ satisfies
$w_{t+1}+w_{t}=\kappa_1$, where $\kappa_1\in\mathbb{R}$ and $\mathbb{R}$ is the set of real numbers.
Note that $\kappa_1$ does not depend on time.
In this case, $w_t$ can be written as $w_t=(-1)^t(w_0-\frac{\kappa_1}{2})+\frac{\kappa_1}{2}$.
Therefore the period of the QW becomes two.
We should remark that
$
 \left[\begin{array}{cc}
  a&b \\c&d
       \end{array}\right] (\,\equiv U\,)
$
is a unitary matrix.
Then we have
\vspace{5mm}

\begin{theorem}
\begin{equation}
 \frac{X_{t}}{t}\,\Rightarrow\,Z_{1},
\end{equation}
where $Z_{1}$ has the density function $f_{1}(x)$ as follows:
\begin{equation}
 f_{1}(x)=f_{K}(x;a)
 \left\{1-\left(|\alpha|^2-|\beta|^2+
 \frac{a\alpha\overline{b\beta}e^{iw_{0}}+\overline{a\alpha}b\beta
 e^{-iw_{0}}}{|a|^2}
\right)x\right\}.
\end{equation}
\end{theorem}

\vspace{5mm}

\begin{proof}

The essential point of this proof is that this case results in the usual walk.
First we see that $U_t$ can be rewritten as
\begin{eqnarray}
 U_t&=&\left[\begin{array}{cc}
       e^{iw_t/2}& 0\\ 0&e^{-iw_t/2}
	    \end{array}\right]
 \left[\begin{array}{cc}
  a& b\\ c&d
       \end{array}\right]
 \left[\begin{array}{cc}
  e^{iw_t/2}& 0\\0 &e^{-iw_t/2}
       \end{array}\right]\nonumber\\[2mm]
 &=&R\left(\frac{w_t}{2}\right) U R\left(\frac{w_t}{2}\right).
\end{eqnarray}
From this, the Fourier transform $\ket{\hat\Psi_{t}(k)}$ can be computed in the following.

\begin{eqnarray}
 \ket{\hat\Psi_{t}(k)}&=&\left\{R(k) R\left(\frac{w_{t-1}}{2}\right) U
 R\left(\frac{w_{t-1}}{2}\right)\right\}
 \left\{R(k) R\left(\frac{w_{t-2}}{2}\right) U
 R\left(\frac{w_{t-2}}{2}\right)\right\}\nonumber\\
 &&\cdots\left\{R(k) R\left(\frac{w_{0}}{2}\right) U
 R\left(\frac{w_{0}}{2}\right)\right\}\ket{\hat\Psi_{0}(k)}\nonumber\\[2mm]
&=&R\left(-\frac{w_t}{2}\right)\left\{R\left(\frac{w_t}{2}\right)R(k) R\left(\frac{w_{t-1}}{2}\right) U\right\}\nonumber\\
&&\times\left\{R\left(\frac{w_{t-1}}{2}\right)R(k)R\left(\frac{w_{t-2}}{2}\right) U\right\}\nonumber\\
 &&\times\cdots\times\left\{R\left(\frac{w_1}{2}\right)R(k) R\left(\frac{w_{0}}{2}\right) U
 \right\}R\left(\frac{w_{0}}{2}\right)\ket{\hat\Psi_{0}(k)}\nonumber\\[2mm]
&=&R\left(-\frac{w_t}{2}\right)\left\{R(k+\kappa_1/2)U\right\}^t R\left(\frac{w_{0}}{2}\right)\ket{\hat\Psi_{0}(k)}.
\end{eqnarray}

Therefore we have
\begin{eqnarray}
 \ket{\psi_t(x)}&=&\int_{0}^{2\pi}\frac{dk}{2\pi}e^{ikx}\ket{\hat\Psi_{t}(k)}
=\int_{\kappa_1/2}^{2\pi+\kappa_1/2}\frac{dk}{2\pi}e^{i(k-\kappa_1/2)x}\ket{\hat\Psi_t(k-\kappa_1/2)}\nonumber\\
&=&e^{-i\kappa_1x/2}R\left(-\frac{w_t}{2}\right)\int_{\kappa_1/2}^{2\pi+\kappa_1/2}\frac{dk}{2\pi}e^{ikx}\left(R(k)U\right)^t \ket{\hat\Psi^R_0(k)},
\end{eqnarray}
where $\ket{\hat\Psi^R_0(k)}=R\left(\frac{w_{0}}{2}\right)\ket{\hat\Psi_{0}(k-\kappa_1/2)}$.
Then the probability distribution is
\begin{eqnarray}
 \lefteqn{P(X_t=x)}&&\nonumber\\
 =&&\left\{e^{i\kappa_1x/2}\left(\int_{\kappa_1/2}^{2\pi+\kappa_1/2}\frac{dk}{2\pi}e^{ikx}\left(R(k)U\right)^t \ket{\hat\Psi^R_0(k)}\right)^{\!\!\ast}R\left(\frac{w_t}{2}\right)\right\}\nonumber\\
 &&\times\left\{e^{-i\kappa_1x/2}R\left(-\frac{w_t}{2}\right)\left(\int_{\kappa_1/2}^{2\pi+\kappa_1/2}\frac{dk}{2\pi}e^{ikx}\left(R(k)U\right)^t \ket{\hat\Psi^R_0(k)}\right)\right\}\nonumber\\
=&&\left|\left|\int_{\kappa_1/2}^{2\pi+\kappa_1/2}\frac{dk}{2\pi}e^{ikx}\hat U(k)^t \ket{\hat\Psi^R_0(k)}\right|\right|^2,
\end{eqnarray}
where $\hat U(k)=R(k)U$.
This implies that Case 1 can be considered as the usual QW with the initial state
$\ket{\hat\Psi^R_0(k)}=R\left(\frac{w_{0}}{2}\right)\ket{\hat\Psi_{0}(k-\kappa_1/2)}$ and the unitary matrix $U$.
Then the initial state becomes
\begin{equation}
 \ket{\hat\Psi^R_0(k)}={}^T[e^{iw_0/2}\alpha,\,e^{-iw_0/2}\beta],
\end{equation}
that is,
\begin{equation}
 \ket{\psi_0(x)}=\left\{\begin{array}{ll}
	       {}^T[e^{iw_0/2}\alpha,\,e^{-iw_0/2}\beta]& (x=0)\\[2mm]
		      {}^T[0,\,0]& (x \neq 0)
		     \end{array}\right..
\end{equation}
Finally, by using the result in Konno \cite{konno_2002_1,konno_2005_1}, we can obtain the desired limit
distribution of this case.
\qed
\end{proof}

\subsection{Case 2}

Next we consider the QW whose dynamics is defined by the following unitary matrix:
\begin{equation}
 U_t=\left[\begin{array}{cc}
      a& be^{iw_t} \\ ce^{-iw_t}& d
	   \end{array}\right].
\end{equation}
Here $w_t\in\mathbb{R}$ satisfies
$w_{t+1}=w_{t}+\kappa_2$, where $\kappa_2\in\mathbb{R}$ does not depend on $t$.
In this case, $w_t$ can be expressed as $w_t=\kappa_2 t+w_0$.
Noting $U_t=R\left(\frac{w_t}{2}\right) U
R\left(-\frac{w_t}{2}\right)$, we get a similar weak limit theorem as Case 1:
\vspace{5mm}

\begin{theorem}
\begin{equation}
 \frac{X_{t}}{t}\,\Rightarrow\,Z_{2},
\end{equation}
where $Z_{2}$ has the density function $f_{2}(x)$ as follows:
\begin{equation}
 f_{2}(x)=f_{K}(x;a)
 \left\{1-\left(|\alpha|^2-|\beta|^2+
 \frac{a\alpha\overline{b\beta}e^{-iw_{0}}+\overline{a\alpha}b\beta
 e^{iw_{0}}}{|a|^2}
\right)x\right\}.
\end{equation}
\end{theorem}

\vspace{5mm}

If $w_t=2\pi t/n\,(n=1,2,\ldots)$, $\left\{U_t\right\}$ becomes an $n$-period sequence.
In particular, when $n=2$ and $a,b,c,d\in\mathbb{R}$, $\left\{U_t\right\}$ is a sequence of two-period orthogonal matrices.
Then Theorem 3 is equivalent to Theorem 1 (i).

\section{Conclusion and Discussion}

In the final section, we draw the conclusion and discuss our two-period walks.
The main result of this paper (Theorem 1) implies that if $\det(H_1H_0)\\>0$ and $\min\left\{|a_0|, |a_1|\right\}=|a_0|$, then the limit distribution of the two-period walk is determined by $H_0$.
On the other hand, if $\det(H_1H_0)>0$ and $\min\left\{|a_0|, |a_1|\right\}=|a_1|$, or $\det(H_1H_0)<0$, then the limit distribution is determined by both $H_0$ and $H_1$.

Here we discuss a physical meaning of our model.
We should remark that the time-dependent two-period QW is equivalent to a position-dependent two-period QW if and only if the probability amplitude of the odd position in the initial state is zero.
In quantum mechanics, the Kronig-Penney model, whose potential on a lattice is periodic, has been extensively investigated, see Kittel \cite{kittel}.
A derivation from the discrete-time QW to the continuous-time QW, which is related to the Schr\"{o}dinger equation, can be obtained by Strauch \cite{strauch}.
Therefore, one of interesting future problems is to clarify a relation between our discrete-time two-period QW and the Kronig-Penney model.

\vspace{5mm}
{\bf Acknowledgment}\\[1mm]
This work was partially supported by the Grant-in-Aid for Scientific Research (C) of Japan Society for
the Promotion of Science (Grant No. 21540118).

\bibliography{main}

\end{document}